\def\year{2022}\relax
\documentclass[letterpaper]{article} 
\usepackage{aaai22}  
\usepackage{times}  
\usepackage{helvet}  
\usepackage{courier}  
\usepackage[hyphens]{url}  
\usepackage{graphicx} 
\usepackage{natbib}  
\usepackage{caption} 
\DeclareCaptionStyle{ruled}{labelfont=normalfont,labelsep=colon,strut=off} 
\usepackage{algorithm}
\usepackage{algorithmic}
\usepackage{amsmath}

\usepackage{newfloat}
\usepackage{listings}
\floatstyle{ruled}
\newfloat{listing}{tb}{lst}{}
\floatname{listing}{Listing}
\nocopyright
\usepackage{subfig}

\title{Opinion Mining from YouTube Captions Using ChatGPT: A Case Study of Street Interviews Polling the 2023 Turkish Elections}
\author {
    Tuğrulcan Elmas,\textsuperscript{\rm 1}
    İlker Gül, \textsuperscript{\rm 2}
}
\affiliations {
    \textsuperscript{\rm 1} Indiana University Bloomington\\
    \textsuperscript{\rm 2} EPFL\\
    telmas@iu.edu, ilker.gul@epfl.ch
}

\usepackage{bibentry}

\begin{document}
\newcommand{\Secref}[1]{Section~\ref{#1}}
\newcommand{\Figref}[1]{Figure~\ref{#1}}
\maketitle
\begin{abstract}

Opinion mining plays a critical role in understanding public sentiment and preferences, particularly in the context of political elections. Traditional polling methods, while useful, can be expensive and less scalable. Social media offers an alternative source of data for opinion mining but presents challenges such as noise, biases, and platform limitations in data collection. In this paper, we propose a novel approach for opinion mining, utilizing YouTube's auto-generated captions from public interviews as a data source, specifically focusing on the 2023 Turkish elections as a case study. We introduce an opinion mining framework using ChatGPT to mass-annotate voting intentions and motivations that represent the stance and frames prior to the election. We report that ChatGPT can predict the preferred candidate with 97\% accuracy and identify the correct voting motivation out of 13 possible choices with 71\% accuracy based on the data collected from 325 interviews. We conclude by discussing the robustness of our approach, accounting for factors such as captions quality, interview length, and channels. This new method will offer a less noisy and cost-effective alternative for opinion mining using social media data. 
\end{abstract}
\section{Introduction}

Opinion mining is a vital method in understanding public sentiment and preferences across various domains, with political elections being a particularly relevant application. Assessing voter intentions and opinions is crucial for predicting election outcomes, shaping campaign strategies, fostering democratic discourse, and promoting informed decision-making by both the electorate and political leaders. Traditional polling methods have been the go-to technique for gauging public opinion; however, they are often expensive, and less scalable. 

The rapid growth of social media has introduced an alternative source for opinion mining, offering a vast and diverse data pool at a lower cost. Despite its advantages, social media data presents its own set of challenges, such as noise, unstructured information, and being limited to users who only (actively) use the platform, which may significantly impact the accuracy of extracted insights. The most popular data source for social media studies was Twitter. This is because Twitter data was easy to collect and process as most of the conversations take place by a short text that can be acquired through the platform's official API~\cite{tufekci2014big}. However, the presence of bots, fake accounts, and other malicious actors raised concerns about the validity of the results~\cite{elmas2023impact}. Additionally, the platform has recently made policy changes to its official API for data provision, which significantly limited data collection. As a result, it is crucial to turn to alternative data sources to reliably mine public opinion.

One of the biggest challenges in using social media data from platforms other than Twitter is the multimodal nature of most of the data. For instance, YouTube data consists of an audiovisual component that can go arbitrarily large. The public opinion presented in this component can be mined through captions. However, this data is unstructured and may be challenging to process as there is no standard video format. Meanwhile, the recent advances in Large Language Models show potential in mining public opinion data from such unstructured data. Moreover, the most recent and advanced LLMs can take multi-modal input, showing that the future LLMs may be able to process entire videos for opinion mining~\cite{gpt4}.

Inspired by these recent significant changes in the data science ecosystem, we propose a new approach for opinion mining and make three key contributions to the field. First, we propose a novel framework to mine public opinions by leveraging YouTube's auto-generated captions from public interviews. We specifically focus on the street interviews conducted for polling the 2023 Turkish elections as a case study due to the popularity of such videos. These interviews consist of a reporter asking random individuals passing by the street about their voting preferences and the motivations behind their choices. Such data offers several advantages over traditional polls and social media data. First, interviews provide a more natural and spontaneous representation of public opinion, as people express their views in real time and without the influence of online echo chambers. Second, the data is considerably cleaner and less noisy than social media data, as it is less susceptible to manipulation by fake accounts and bots. Finally, the costs associated with gathering and processing such interview data are significantly lower than those of traditional polling methods.

Our second contribution is the introduction of an opinion-mining framework that employs YouTube data. We present a case study where we employed ChatGPT to mass-annotate opinions that are represented by YouTube captions to understand public opinion prior to the 2023 Turkish general elections. Our framework shows the potential to provide insights about public opinions in an efficient manner, making YouTube data a potential alternative to Twitter data.

Our last contribution is presenting a case study that shows ChatGPT's capability to detect stance (i.e., candidate preference) and frames (i.e., motivations behind choosing a particular candidate) and mass annotate YouTube captions data in a scalable, efficient, and effective analysis of voter intentions and opinions. We do this by creating a ground truth dataset, annotating the captions of 325 interviews from 10 videos published by 3 different channels. We then evaluate the performance of our framework and report that it can predict the preferred candidate with 97\% accuracy and predicts the correct motivation out of 13 possible choices with 71\% accuracy. However, it cannot identify all the respondents using only the captions in plain text form. We also provide a robustness analysis to account for the effect of captions quality, interview length, and channels.

\section{Related Work}

We contribute to opinion mining by introducing a novel framework, testing ChatGPT for a new task, and employing Youtube captions for election polling. We now provide the related work in those areas and highlight our contributions.  

\subsection{Opinion Mining}

The objective of opinion mining is to develop a knowledge base that organizes online opinions in a more coherent and structured form~\cite{sobkowicz2012opinion}. To achieve this, researchers usually collect data that represent opinions and annotate them according to some aspect of them such as sentiment, stance, or topics, which are standalone tasks on their own. 

Sentiment analysis aims to determine the direction of the emotions in the data (positive, negative, or neutral). Meanwhile, stance detection aims to determine the position (e.g., favor, against, or neutral) of a given data concerning a target topic or entity. Past work proposed machine learning and deep learning-based approaches to detect sentiment~\cite{giachanou2016like} and stance of the text~\cite{mohammad2016semeval, augenstein2016stance, baly2018integrating}. A related task is to predict the political orientation of a text or author which may be news articles~\cite{zhou2011classifying} or social media users~\cite{wong2016quantifying}. This task may also use contextual information such as users' social network~\cite{barbera2015birds, elmas2020can}. In some cases, the task may need a more detailed description of the data. Such tasks can be formulated as detection of the topics~\cite{wartena2008topic} or frames~\cite{mendelsohn2021modeling}. In our framework, we tackle the preferred candidate of the respondent and how they frame the candidate, which are similar to stance and frame detection.

\subsection{Using Large Language Models for Data Annotation}

The opinion mining tasks usually turn to human annotation to build ground truth dataset. Then they propose machine learning models to mass annotate the remaining dataset for free. However, large language models may facilitate this process. One of the most recent LLMs, ChatGPT, which is open to the public, was reported to be useful on tasks such as translation~\cite{jiao2023chatgpt}, genre identification~\cite{kuzman2023chatgpt}, hate speech detection~\cite{huang2023chatgpt}. It also shows promising results in data annotation. For instance, Zhong et al.~\citeyear{zhong2023can} reported that ChatGPT outperforms base and large BERT and Roberta models on inference tasks and achieves comparable performance in sentiment analysis and question-answering tasks. Similarly, Gilardi et al.~\citeyear{gilardi2023chatgpt} suggest that ChatGPT outperforms crowd workers in Amazon MTurk in zero-shot accuracy in four tasks: data relevance, stance, general and policy frame detection. We also tested ChatGPT which employs GPT-4 in predicting the candidate preference and motivation detection and report promising results. 

Chu et al.~\citeyear{chu2023language} suggested that ChatGPT can predict public opinions and thus, has the potential to supplement polls and forecast public opinions. This work is one of the first applications towards this goal in which we suggest open-source interviews to supplement traditional polls and analyze a sample of them extensively.

\subsection{Elections Polling using Social Media Data}

Traditional polls are conducted by selecting individuals from the population at random, and then interviewing them in person or through a phone call. As those techniques are expensive, researchers turn to online polls that do not necessarily require directly contacting the respondents. Such polls may not be representative of the population as internet usage differs by population. However, such polls can be adjusted through statistical methods to make the results of the poll better represent the population. For instance, Wang et al~\citeyear{wang2015forecasting} showed that it is possible to forecast polls by adjusting the poll results that are conducted through the Xbox gaming platform which are biased towards younger male population. 

Past studies proposed social media data as an alternative to data acquired by traditional polling. Bovet et al.~\citeyear{bovet2018validation} suggested that opinion trends on Twitter followed the aggregated NYT polls, suggesting that social media can be predictive of opinion trends and election results to some extent. Liu et al.~\citeyear{liu2021can} introduced a model that combines election forecasting models from political science and Twitter sentiment analysis to predict elections. Meanwhile, Brito et al.~\citeyear{dos2020predicting} propose a model that combines both Facebook, Twitter, and Instagram data to predict elections. Those works showed the potential of social media in nowcasting opinion trends or forecasting election results to some extent. 

While collecting social media data is cheaper and easier than conducting polls, it has its drawbacks. Social media is noisy: bots and trolls actively use social media to propagate their narratives to influence the public~\cite{elmas2022characterizing}. Additionally, it is sparse~\cite{saha2021person}, and people may not provide their opinion on a subject even if they have one. A handful of users' opinions may go viral but may not necessarily represent the public opinion~\cite{elmas2023measuring}. The recent changes in Twitter's data access policy also significantly limited the researchers' ability to collect social media data. This makes us turn to alternative data providers. In light of these factors, we introduce a hybrid approach that uses the \textit{data} of public \textit{in-person} interviews. This approach facilitates large-scale analysis while preserving the advantage of employing data collected by in-person interviews. The data we use in our case study are from street interviews that are more likely to be representative than social media platforms like Twitter although it may still not be perfectly representative due to other factors. However, since those are from public videos, they can be adjusted according to demographics to make the sample more representative. Moreover, the data is from YouTube which is a good and free alternative to Twitter.

Our approach combines the pros of both approaches. The street interviews on YouTube is conducted the same way as traditional polling. The only difference is that it is limited to crowded places in the cities and towns versus people who accepted the call or the interviews. Meanwhile, it uses the advantage of social media data; it is widely conducted in Turkey and provides large-scale analysis. The interviews are made by real people, contrary to Twitter where fake or bot accounts thrive. Caption data is free, does not require API access, and imposes fewer ethical risks since people who speak are invisible in the data (unless they explicitly disclose their identity, which is not the case with street interviews).

This work is also the first to use YouTube captions data for mining public opinion to the best of our knowledge. YouTube data is understudied and there is only a few studies using captions. Jagtap et al.~\citeyear{jagtap2021misinformation} employed captions data to detect misinformation in the videos. Dinkov et al.~\citeyear{dinkov2019predicting} employed captions to predict the channel's political ideology. This work will contribute to the field of computational social science by employing alternative data sources such as YouTube data.

\section{Opinion Mining Framework}
\begin{figure*}[!htb]
    \centering
    \includegraphics[width=\linewidth]{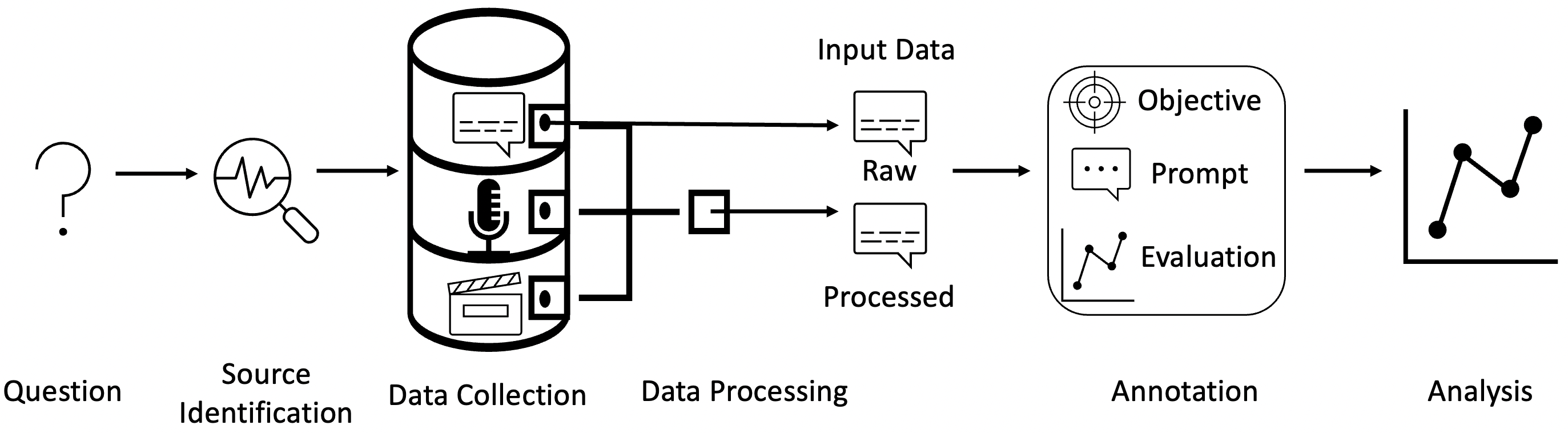}
    \caption{The overview of the annotation framework that is compatible with social media data including YouTube.}
    \label{fig:pipeline}
\end{figure*}

We propose an opinion mining framework that is compatible with YouTube captions data. ~\Figref{fig:pipeline} shows its overview. Our contribution is on the annotation and the analysis part while the fine-tuning and automating of the complete pipeline is left to future work due to current limitations with working with YouTube data and ChatGPT API. 

We now describe each component of the pipeline and provide the methodological details by using our case study as the primary example. We describe the experimental results in the next section. 

\subsubsection{Question:} This part outlines the opinion mining problem that the framework aims to address. It helps define the focus and scope of the study and guides the development of the methodology. In our case, the questions are ``What is the preferred election candidate of the public?" and ``What are the motivations behind their choices?". We can answer the former using a quaternary political stance classification of each respondent to account for the three candidates and the ``other" option. The latter question can be formulated as detecting the topic or the concept that best describes the motivation that is disclosed by the respondent, or how they ``frame" the candidate. We choose the former option and prepare our annotation tasks accordingly, which we describe in the respective component. 

\subsubsection{Source Identification:} In this part we select the appropriate data sources to answer the question. First, we select the online platform to collect the data. Then we select the source within the platform where the relevant data may be, i.e., specific sources, channels, users, or the data that are tagged by, or contain the keywords that are relevant to the question. Such sources or keywords may be selected manually or by identifying relevant keywords in an automated manner~\cite{vijaya2022graph}. On Twitter, researchers studying elections usually use candidate names (e.g., ``Donald Trump") and their buzzwords (e.g., \#BuildTheWall)~\cite{bovet2018validation}. The data can be further limited to users according to their relevance to the question (e.g., limiting only to users in the country where the elections take place, filtering out the bots). In our case, we search for the videos containing the keyword ``Election Poll" (\textit{Seçim Anketi}) and limit the data to the videos from the channels whose main purpose is to conduct street interviews. 

\subsubsection{Data Collection:} In this part, we collect data from the identified sources. On Twitter, this translates to identifying the appropriate API endpoint, and optionally selecting a timeframe, languages, maximum number of results, etc. The data is usually in short-text form but may also contain images and videos. On YouTube, this translates to collecting the complete video or its description, captions, and audio from the identified sources. We specifically focus on the captions data in this study.

\subsubsection{Data Processing:} In this part we clean, preprocess, and transform the collected data into a suitable format for annotation, and analysis. On Twitter, this may be removing spam, expanding URLs, and building Twitter threads (as they are made up of multiple data points). In our study, we process the captions to account for the different people speaking in the same video to yield more reliable results. We both experiment with raw and processed captions to point out the difference in the performance. We explain the processing steps we employed in detail in the next section.

\subsubsection{Annotation:} The annotation process has three components. The first is to determine the annotation objective which focuses on the specific aspect of the data that annotators need to examine and label. We determine the desired input and output in this component. Then we prepare the prompt (i.e., annotation question) to show to the annotators which will guide them in the annotation process and ensure that the annotators understand the goal. It may include definitions, examples, and potential edge cases to ensure a standardized approach. The annotators can be individuals recruited from crowdsourcers or experts. In the third component, we evaluate the annotation quality through computing annotation agreement. In the case of low agreement, we repeat the whole process. In our approach, we use ChatGPT which employs the GPT-4 model to annotate the captions with respect to the preferred candidate and the motivations of the respondents. We test its performance against human annotation. Using ChatGPT is advantageous to traditional annotation as it is cheaper and more efficient. It is also easier to introduce changes to the process. For instance, it is easier to tune the prompt to get the best results while it is expensive to change the annotation question shown to the annotators once they do the annotation.

\subsubsection{Analysis:} After the annotation, we provide a quantitative analysis in which we describe the phenomena we are interested in. In our case, this is counting and reporting the annotation results, which are the respondents' preferences of candidates and motivations. This is similar to what has been done on Twitter or other social media studies.

\section{Experiments}

\begin{figure*}[!htb]
\includegraphics[width=\linewidth]{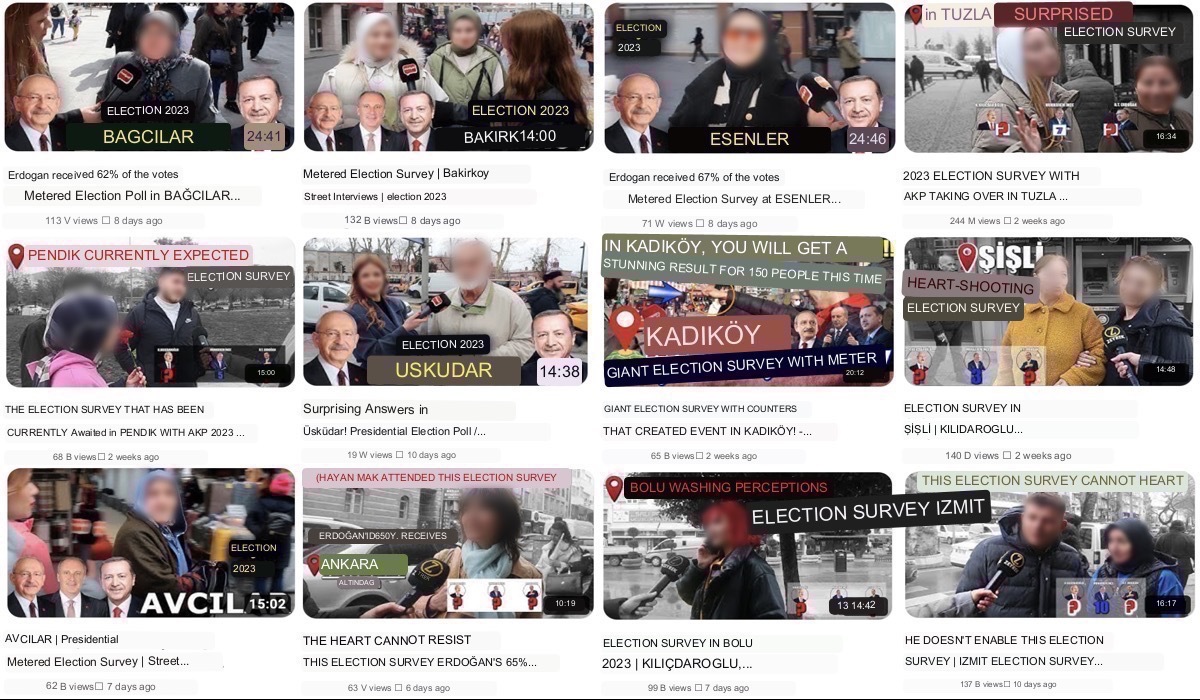}
\caption{The videos used in the study. All are from March 2023, 2 months before the election, and have thousands of views.}
\label{fig:videos}
\end{figure*}

\begin{figure}[!htb]
\subfloat{\includegraphics[width=0.9\linewidth]{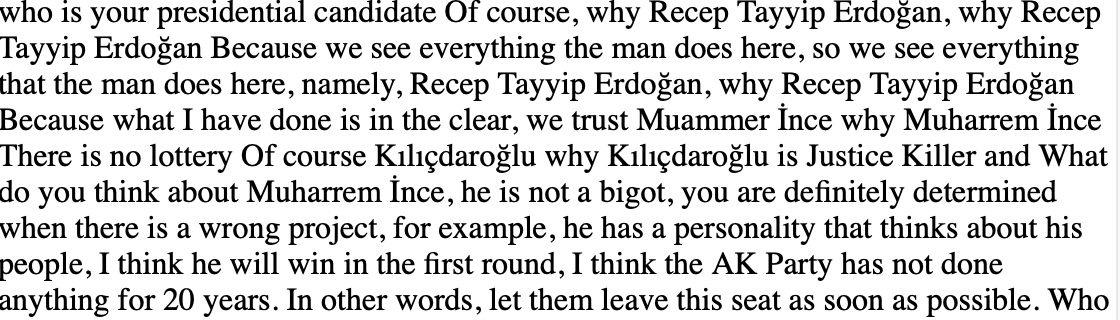}}
\newline
\subfloat{\includegraphics[width=0.7\linewidth]{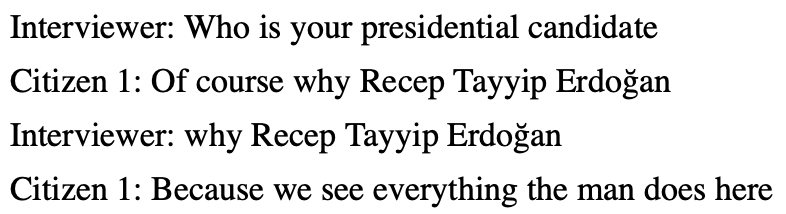}}

\caption{Plain-text format (up) and the processed format (down) in which we distinguished the speakers. The second sentence has a speech-to-text error which we did not correct.}
\label{fig:plaintext}
\end{figure}

In this section, we follow the opinion mining framework we outlined previously and adapt them to our case study in which we analyze the voting intentions and motivations of the public before the 2023 Turkish general election. We present the methodological and experimental details in data collection, captions processing, building ground truth, and evaluation metrics. 

\subsection{Data Collection} We search for street interviews using the query ``Seçim Anketi" (Election Poll) and manually identify the videos which feature an interviewer asking random people's opinions on the street. As such videos are very popular and easily hit 100,000 views after a day since their upload, they rank high in the search results, and thus, are very easy to identify. A caveat is that the channels may cherry-pick the interviews and be biased toward certain respondents and candidates. To ensure this does not bias the results, we picked 12 videos from three different channels to ensure diversity. The channels are Medyali, Zeyrek and Halk Ekranı. We specifically chose these channels as they provide the results of the survey which yield easier and more reliable annotation. Additionally, they state that their videos are unedited, which we can confirm. All videos are from different locations. 9 videos are from different districts of Istanbul and the 3 videos are from other provinces. Figure~\ref{fig:videos} shows the video thumbnails, and Table 1 shows the districts. We downloaded their captions using captionsgrabber.com.

\subsection{Captions Processing} The captions we download are in plain-text format. As each video features several interviews (roughly 2-3 interviews per minute), there are multiple people speaking. The plaintext format does not differentiate speakers and there are no other formats that provide such information. In our preliminary experiments, we observe that it is not easy for humans or for ChatGPT to distinguish people by only looking at plain text. Thus, we need a new subtitle format that distinguishes the speakers. We can produce data in this format using the audio data and tagging the speakers. The format we choose provides ids to the speakers and transcribes what they say without correcting the speech-to-text mistakes. We manually processed the captions to follow this format for testing purposes. We leave an automatic approach to mass-process captions to future work.

We refer to the captions in the plain-text format as \textbf{raw} data and the captions in the identity-tagged format as the \textbf{processed} data. Figure~\ref{fig:plaintext} shows an example of the two formats. 

\subsection{Ground Truth} The main objective of the experiment is to test GPT-4's capability to annotate the captions so that we can use it to mine public opinions. For this purpose, we annotate the captions by ourselves and then reannotate them using the GPT-4 to evaluate their performance. We have two annotation tasks: the annotation of the preferred candidate and the annotation of the motivation. 

\subsubsection{Preferred Candidate:} In each video, the respondents are asked their preference for one of the candidates, Recep Tayyip Erdoğan (RTE), Kemal Kılıçdaroğlu (KK), Muharrem İnce (Ince). If they do not decide on one of those choices, they are labeled as ``Other or Undecided". ``Other" consists of the candidates that were not official at the time (Cem Uzan, Sinan Oğan, Selahattin Demirtaş). All the interviews in our dataset considered those four classes. The only exception is the one shot in Bağcılar in which İnce is also considered as ``Other". Since all the videos self-state their respondents' preferences (i.e., they provide a counter at the bottom of the video), we use their annotations as the ground truth and do not annotate by ourselves. 

\subsubsection{Motivation:} The motivation is why the respondent prefers the candidate. They can be very diverse as the interviews are conducted by employing the open-ended question ``Why?". Thus, predicting them and evaluating the predictions are tricky. To overcome this challenge, we annotate voters' annotations through a single concept, which is the aspect of the candidate the respondent highlight. It is also called a frame in the media literature~\cite{mendelsohn2021modeling}. We can either predetermine the concepts by annotating the data by ourselves or ask the annotator (ChatGPT) to do that. Predetermining the concepts may limit the annotators' understanding. In our preliminary experiments, we also find that relying only on ChatGPT introduces the problem of granularity and consistency. For instance, ChatGPT yields two different concepts for similar motivations, ``leadership" and ``being a world-class leader" which can be merged into one. The leadership is represented by ``charisma" in another video. To account for this problem, we employed a mixed approach. We used 20\% of the data to predetermine the concepts by simultaneously using expert annotations and ChatGPT annotations. In other first, we first annotated the data using ChatGPT, which revealed 28 different concepts from 60 interviews. Then one author of this paper looked at those annotations and merged the concepts that share the same root (leader, leadership) or are semantically similar. He also added four other concepts that are not mentioned by ChatGPT but the annotator feel should have been mentioned given the data. The other author examined and verified those annotations. As a result, we determine the list to make up 12 concepts and ``other". We did not correct ChatGPT's annotations in order not to leak test data to the model. In other words, ChatGPT's annotations are only used to help the annotators predetermine the list of concepts. 

\begin{figure}
    \centering
    \includegraphics[width=0.8\linewidth]{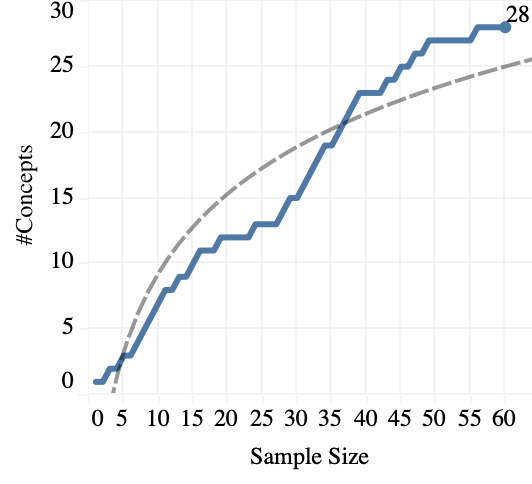}
    \caption{The number of new concepts yielded by ChatGPT per sample size.}
    \label{fig:annotation_robustness}
\end{figure}

We also provide a robustness analysis by showing that the number of new concepts decreases by sample size and remain stable after 45 interviews in Figure~\ref{fig:annotation_robustness}. This shows that in future applications where the classification problem is multiclass and the classes are not determined, a sample of data can be employed to decide them on the fly. We can determine the sample size by looking at the number of new classes introduced by each new sample and this process when the number of new classes stays stable. 

The final list of concepts is listed in Table 2. This list covers 84\% of all the motivations in the whole data. We provide the annotation question in the next section as the same annotation question was used to predetermine the concepts and to evaluate ChatGPT. 

\subsection{Annotation}

We prepared the annotation question as if it will be presented to human authors. We provide the data, describe the form of the data, and instructions on how to handle corner cases (e.g., multiple respondents.). We then provide the data and the list of concepts. We then list our questions.  ChatGPT was given the Turkish version of the following prompt (i.e., annotation question).

\textit{Below, you can find the text of the interviews between the reporter and the citizens in the YouTube video for the election survey about whom they will vote in the elections, in timestamped form, and the list of concepts that will help us to conceptualize the reasons for the vote.}

\textit{Important Details:}

\textit{If more than one citizen is being interviewed at the same time, citizens are named as Citizen 1, Citizen 2.}

\textit{The answers given to the question ``Why not ... ?" do not indicate the votes of the citizens.}

\textit{Captions: ...}

\textit{List of concepts: ...}

\textit{Questions:}

\textit{Question 1: How many citizens were interviewed in total?}

\textit{Question 2: Provide each citizen's preferred candidate, the reason for voting, and the one-word concept summarizing the reason by choosing the most suitable concept from the list of concepts, in .csv format. If they didn't give a reason, please don't give a reason either. Choose ``other" if you don't think there is a suitable concept from the list of concepts. If the citizen is undecided, mark the reason and the concept as undecided.}

\textit{Answers:}

\textit{Answer 1:}

\textit{Answer 2:}

We prepared the important details part based on our preliminary experiments. Precisely, we observed that ChatGPT performed better if we explicitly state that multiple citizens may be interviewed simultaneously (as some people join ongoing interviews while passing by the street) and some citizens were asked about their reasons for not preferring another candidate. We also observed that ChatGPT does not provide predictions for all the respondents if it surpasses the output limit. In these cases, we instructed ChatGPT to continue predicting the remaining respondents. In order to check if it can do this, we explicitly ask it to provide the number of respondents in the beginning so that we can check if the list of its predictions matches the number of respondents. We also request the respondents' motivations for our records, without limiting them to predetermined concepts. Finally, we explicitly ask it to provide the output in CSV format to facilitate analysis. 

In our preliminary analysis, we observed that GPT 3.5 cannot solve this task reliably. Thus, we used GPT-4 through ChatGPT, as the API was not available at that time.

\subsection{Evaluation} 

We have two classification problems to evaluate: candidate and motivation predictions. The former has four classes while the latter has thirteen, making them multi-class classification problems. For simplicity, we use the one versus the rest approach. In our problem, we consider each candidate and concept equal. Therefore, we use micro-averaged precision and recall which gives equal weight to all instances~\cite{sokolova2009systematic}. Although this may favor big classes, we do not have such a concern in this problem. We consider each correct classification as true positive $tp$, and incorrect classification as false positive $fp$. Not correctly classifying the preference and the motivation of a respondent is considered false negative $fn$. Hence, precision is $tp / (tp + fp)$, and recall is $tp / (tp + fn)$. Thus, both precision and recall are equal to accuracy if the classifier makes a prediction for each respondent. However, in some cases, the classifier cannot identify the response of a respondent and thus, does not yield a classification result. We do not consider such cases as false positives. Thus, the precision is higher than recall when the classifier does not yield predictions. This occurs when ChatGPT cannot distinguish different respondents from the captions in the plaintext format.

\section{Results}

\subsection{Overall Results}

We show the channel-wise and combined results in Table 1 and concept-wise results in Table 2. 

On 325 respondents, we report 0.97 precision and recall for candidate prediction and 0.70 for concept classification, if the captions are processed, i.e., the speakers are provided to ChatGPT. This means that ChatGPT may reliably detect people's stances and the frames they mention from YouTube data.

\begin{table*}[htbp]
\caption{All and the channel-wise results of the candidate preference and the concept classification (shown by the prefix Co).}
\label{tab:channels}
\resizebox{\linewidth}{!}{
\centering
\begin{tabular}{|l|l|l|l|l|l|l|l|l|l|l|l|l|}
Location &  All &  RTE &   KK & Ince &  Prec &  Prec* &  Rec &  Rec* &  CoP. &  CoP.* &  CoR. &  CoR.* \\
\hline
Bakırköy &   30 & 23\% & 47\% & 20\% &  0.89 &   0.97 & 0.53 &  0.97 &  0.39 &   0.73 &  0.23 &   0.73 \\
Bağcılar &   26 & 35\% & 35\% &  0\% &  0.95 &   1.00 & 0.69 &  0.96 &  0.42 &   0.72 &  0.31 &   0.69 \\
 Esenler &   28 & 50\% & 39\% &  7\% &  0.94 &   0.96 & 0.57 &  0.96 &  0.59 &   0.71 &  0.36 &   0.71 \\
   Tuzla &   32 & 16\% & 47\% & 19\% &  0.95 &   1.00 & 0.62 &  1.00 &  0.43 &   0.59 &  0.28 &   0.59 \\
  Pendik &   23 & 43\% & 22\% & 17\% &  0.94 &   1.00 & 0.65 &  1.00 &  0.44 &   0.70 &  0.30 &   0.70 \\
 Üsküdar &   21 & 38\% & 43\% &  5\% &  0.91 &   0.95 & 0.48 &  0.90 &  0.55 &   0.80 &  0.29 &   0.76 \\
 Kadıköy &   29 & 10\% & 59\% & 21\% &  0.86 &   0.93 & 0.21 &  0.93 &  0.50 &   0.83 &  0.10 &   0.83 \\
   Şişli &   21 & 24\% & 62\% & 10\% &  0.80 &   0.90 & 0.38 &  0.90 &  0.30 &   0.57 &  0.14 &   0.57 \\
 Avcılar &   22 & 23\% & 32\% & 32\% &  0.90 &   0.95 & 0.41 &  0.95 &  0.50 &   0.86 &  0.23 &   0.86 \\
  Ankara &   31 & 42\% & 28\% &  6\% &  0.89 &   1.00 & 0.55 &  1.00 &  0.42 &   0.65 &  0.26 &   0.65 \\
    Bolu &   29 & 31\% & 38\% & 17\% &  0.81 &   0.97 & 0.45 &  0.97 &  0.62 &   0.76 &  0.34 &   0.76 \\
 Kocaeli &   33 & 45\% & 36\% & 15\% &  0.93 &   0.97 & 0.42 &  0.97 &  0.50 &   0.64 &  0.21 &   0.64 \\
 \hline
     All &  325 & 32\% & 41\% & 14\% &  0.91 &   0.97 & 0.50 &  0.96 &  0.47 &   0.71 &  0.26 &   0.70 \\

\end{tabular}
}
\end{table*}

\subsubsection{The Effect of the Subtitle Processing:} The results worsen substantially if the subtitles are not processed and the speaker information is not provided to ChatGPT. We observe that the recall of the classifying preference decreases from 0.96 to 0.5 overall and concept classification from 0.7 to 0.26. Thus, speaker information is crucial to get reliable results, especially in a study like ours which features multiple respondents in a single video.

\subsubsection{Cross-video performance}: We observe that the results are not consistent across videos. This may be because the video quality is not uniform, e.g., the lowest performance is Kadıköy in which the sound quality is low and the video is shot near the Kadıköy pier where there is a strong wind blowing from the sea. This creates background noise and may be affecting the performance of the speech-to-text. The video format is not uniform across channels as well; some channels keep the interviews very short while others conduct longer interviews. This emphasizes the use of quality data in such a study. 

\subsubsection{Concept-wise performance:} We observe that voters of Erdoğan usually emphasize Leadership, Development, and Stability while voters of Kılıçdaroğlu and İnce to some extent emphasize the need for a Change, Economy, and Honesty. ChatGPT does well in most of the concepts except for Persistence and Leadership. This is because the way these concepts are presented varies across candidates, which should be accounted for.

\begin{table}[htbp]
\label{tab:concepts}
\caption{The results of the concept predictions and their distribution across candidates.}
\resizebox{\linewidth}{!}{
\centering
\begin{tabular}{|l|l|l|l|l|l|l|l|l|l|}
Concept & Explanation & RTE & KK & Ince & Prec & Prec* & Rec & Rec* \\
\hline

Leadership & He is a good or charismatic leader & 34 & 8 & 2 & 0.39 & 0.57 & 0.24 & 0.53 \\
Change & The country needs a change & 0 & 27 & 6 & 0.62 & 0.85 & 0.47 & 0.85 \\
Economy & The economy is bad, people are poor & 1 & 14 & 3 & 0.69 & 0.77 & 0.50 & 0.72 \\
Development & He developed the country & 17 & 0 & 0 & 0.67 & 0.75 & 0.47 & 0.71 \\
Honesty & He does not lie and is not corrupt & 1 & 10 & 2 & 0.57 & 0.57 & 0.31 & 0.77 \\
Stability & The country needs politically stable & 10 & 0 & 0 & 0.40 & 0.80 & 0.20 & 0.70 \\
Intimacy & People love him & 0 & 4 & 4 & 0.67 & 1.00 & 0.25 & 1.00 \\
Reliable & You can count on him & 3 & 3 & 1 & 0.33 & 0.67 & 0.29 & 0.71 \\
Persistance & He can withstand the adversities & 2 & 1 & 3 & 0.20 & 0.40 & 0.17 & 0.33 \\
Had Enough & Had enough of the current system & 0 & 5 & 0 & 0.67 & 0.67 & 0.40 & 0.60 \\
Justice & He will be fair to everyone & 0 & 3 & 0 & 1.00 & 1.00 & 1.00 & 1.00 \\
Faith & He is religious & 2 & 0 & 0 & 1.00 & 1.00 & 0.50 & 1.00 \\

\end{tabular}
}
\end{table}

\subsubsection{Interview Length:} 
\begin{figure*}[!htb]
\subfloat{\includegraphics[width=0.30\linewidth]{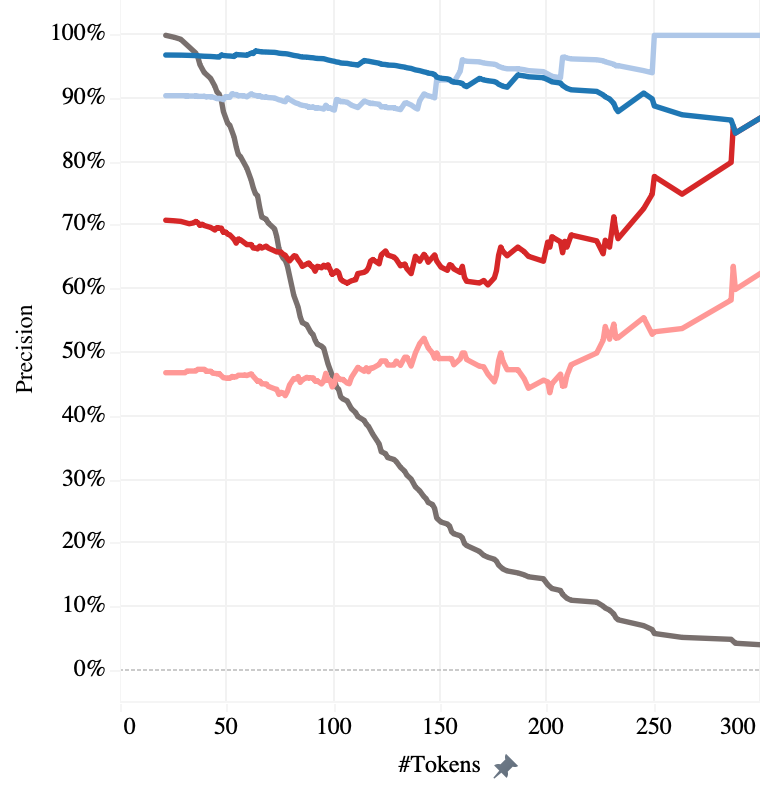}}
\subfloat{\includegraphics[width=0.30\linewidth]{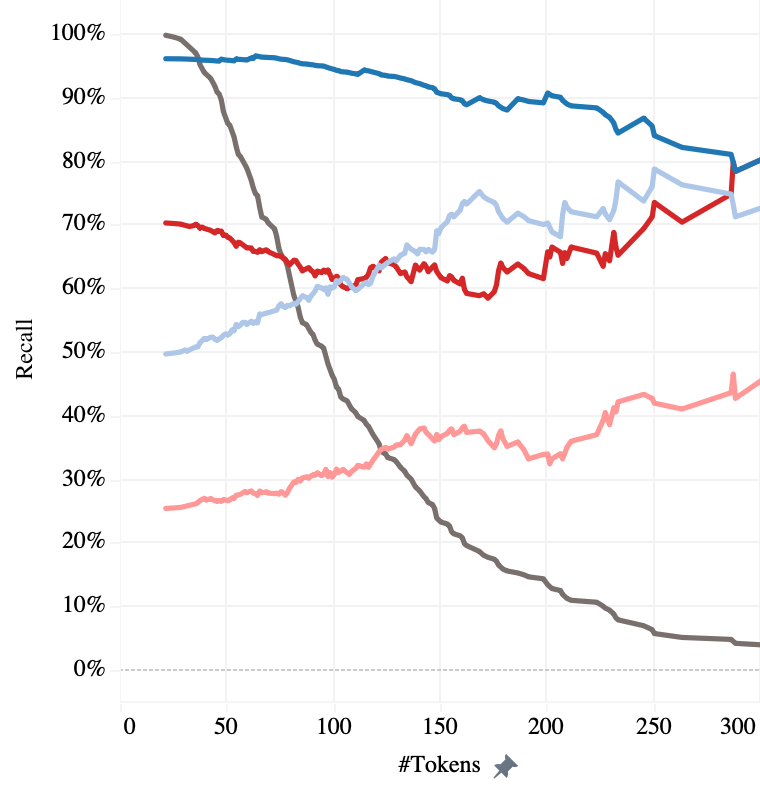}}
\subfloat{\includegraphics[width=0.30\linewidth]{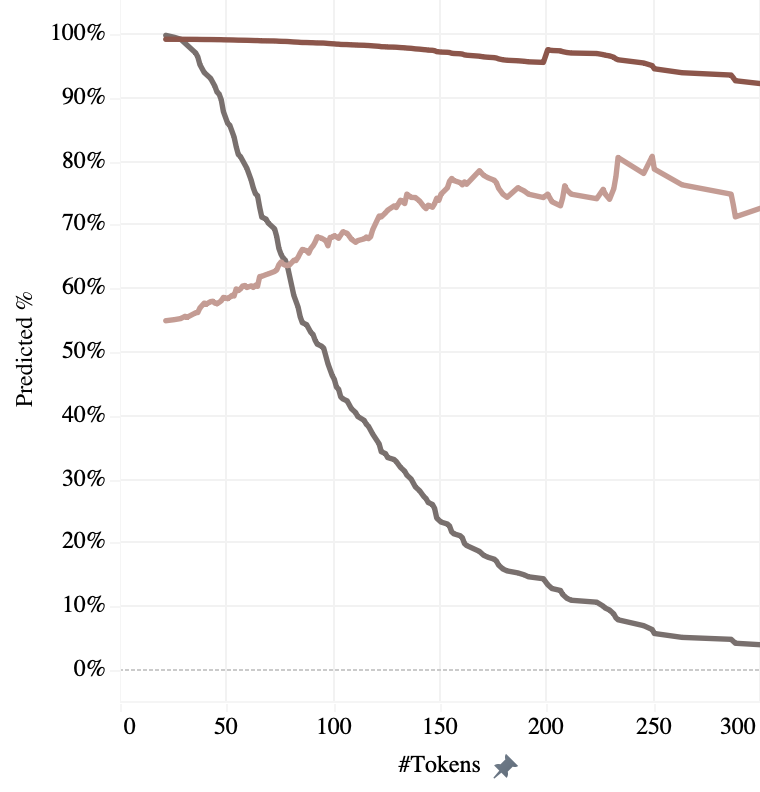}}
\raisebox{0.9\height}{\begin{minipage}[t]{0.15\linewidth}
    \includegraphics[width=\linewidth]{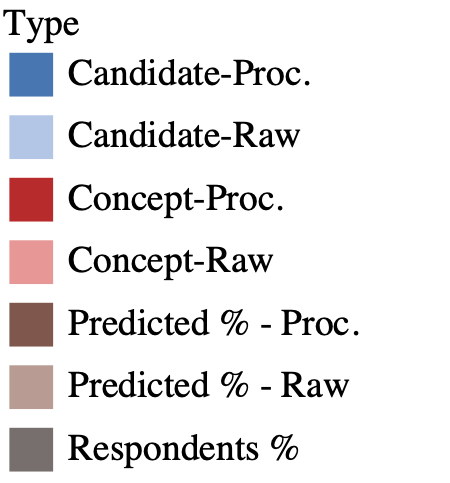}
\end{minipage}}
\caption{Precision (left), Recall (middle), and the percentage of respondents the classifier yield a prediction (right), with respect to token count of the interview.}
\label{fig:tokencount}
\end{figure*}

We observe that interview lengths vary and affect the prediction. To analyze this, we computed the token count of each interview. We then rebuild the dataset by imposing a minimum token count and reevaluating the predictions. \Figref{fig:tokencount} shows the results. We observe that the prediction performance on the raw captions improves with the token count. This means that when the interviews are longer, ChatGPT can better distinguish the respondents and predict their preferences better. Thus, in another case study where the interviews are usually longer, ChatGPT may provide much better results. The concept classification improves both for raw and processed captions with interview length as the respondents explain their motivations over a longer period which provides more input for ChatGPT to detect concepts. Interestingly, the performance in candidate prediction slightly worsens with the interview length for processed captions. This is because, in long interviews, some respondents talk extensively about other candidates which confuses ChatGPT even though we mitigated this problem by explicitly stating that they should not consider the answer to ``Why not X?" as a voting preference. We also observe that the number of respondents decreases significantly for marginal gains in precision and recall. Thus, it may be a better strategy not to filter the data according to interview length, especially if processed captions are employed.

\subsubsection{Language:} We only used the videos in the Turkish language in this study as election polling through street views is popular to a large scale only in Turkey to the best of our knowledge. However, we observe that YouTube's text-to-speech is poor when compared to English. The results may improve if the framework is employed for English videos but may worsen for low-resource languages. We leave the analysis of the effect of language on the results to future work.

\section{Implications on Journalism}


In the traditional journalism model, media agencies employ interviewers to conduct field interviews and then use the responses to prepare news reports. However, the rise of social media has provided alternative ways to mine public opinion such as by analyzing tweets. Nevertheless, social media suffers from low public penetration and demographic imbalances. For example, elderly people are less likely to use social media. Additionally, social media contains bots and other politically motivated individuals that do not represent the public reliably. As a result, media agencies have adopted a hybrid approach, using both traditional field interviews and social media sources.

However, our study shows that publicly available internet interviews can be analyzed and reported on, implying that interviews can be open-sourced and media agencies can outsource interviews to independent interviewers. These interviewers can make videos for their YouTube channels, and newspapers can analyze them at scale using our approach. Thus, our study contributes to open-source and automated journalism, using social media data.

\section{Conclusion}
In this work we introduced an opinion-mining pipeline that employs ChatGPT for mass annotation of interview data, which contributes to the fields of opinion mining, political science, and public opinion research. We believe that these contributions will pave the way for future research and practical applications in understanding public sentiment more accurately and efficiently.

By analyzing the captions of these street interviews, we also extract valuable insights into voter intentions and opinions, which can contribute to a more accurate understanding of the political landscape in the 2023 Turkish elections. Our study demonstrates the potential of this innovative data source for opinion mining and highlights the importance of exploring alternative avenues to better capture a public sentiment in the digital age.

\section{Future Work}

In this section, we discuss potential avenues for future research and improvements to build upon the current findings.

\subsubsection{Identifying Videos:} In this work, we identified the videos manually as we do not work with a massive sample. However, in future work, we will increase the sample size by introducing more channels and videos. Thus, we will also implement video retrieval to the source identification component in the pipeline. 

\subsubsection{Reducing Costs:} In this work, we used the ChatGPT interface which was unlimited given that the user pays the monthly subscription fee. However, the API limitations are different and billed per prompt and output token count. Thus, it is crucial to tune the prompts to reduce the cost. We leave this task to future work. 

\subsubsection{Adjusting for the Demographical Bias:} While people walking on a crowded street in the city center may be closer to random sampling than using social media data, it has still subject to biases that we have to acknowledge and mitigate in future work. First, the streets the interviews are located may be preferred by some demographic groups over others. Secondly, some interviews may be staged, although it is not likely that all of them will be as staged interviews are more likely to be more costly. Thirdly, even if they are not staged, interviewers may be more biased to interview a certain demographic group, e.g., some internet users claimed that some channels were biased to interview young people to inflate Ince's votes. Those street interviews can still be adjusted by demographics, which we leave to future work. 

\section{Ethical Disclosure} 

In this work, we only used the captions which were provided by YouTube publicly. We did not use any data that exposes the identity of the respondents or any other people. The respondents have given their consent to be filmed and published explicitly.

\bibliography{aaai22.bib}

\end{document}